\catcode`\@=11					



\font\fiverm=cmr5				
\font\fivemi=cmmi5				
\font\fivesy=cmsy5				
\font\fivebf=cmbx5				

\skewchar\fivemi='177
\skewchar\fivesy='60


\font\sixrm=cmr6				
\font\sixi=cmmi6				
\font\sixsy=cmsy6				
\font\sixbf=cmbx6				

\skewchar\sixi='177
\skewchar\sixsy='60


\font\sevenrm=cmr7				
\font\seveni=cmmi7				
\font\sevensy=cmsy7				
\font\sevenit=cmti7				
\font\sevenbf=cmbx7				

\skewchar\seveni='177
\skewchar\sevensy='60


\font\eightrm=cmr8				
\font\eighti=cmmi8				
\font\eightsy=cmsy8				
\font\eightit=cmti8				
\font\eightbf=cmbx8				

\skewchar\eighti='177
\skewchar\eightsy='60


\font\ninei=cmmi9
\font\ninesy=cmsy9

\skewchar\ninei='177
\skewchar\ninesy='60


\font\tenrm=cmr10				
\font\teni=cmmi10				
\font\tensy=cmsy10				
\font\tenex=cmex10				
\font\tenit=cmti10				
\font\tensl=cmsl10				
\font\tenbf=cmbx10				
\font\tentt=cmtt10				
\font\tenss=cmss10				
\font\tensc=cmcsc10				
\font\tenbi=cmmib10				

\skewchar\teni='177
\skewchar\tenbi='177
\skewchar\tensy='60

\def\tenpoint{\ifmmode\err@badsizechange\else
	\textfont0=\tenrm \scriptfont0=\sevenrm \scriptscriptfont0=\fiverm
	\textfont1=\teni  \scriptfont1=\seveni  \scriptscriptfont1=\fivemi
	\textfont2=\tensy \scriptfont2=\sevensy \scriptscriptfont2=\fivesy
	\textfont3=\tenex \scriptfont3=\tenex   \scriptscriptfont3=\tenex
	\textfont4=\tenit \scriptfont4=\sevenit \scriptscriptfont4=\sevenit
	\textfont5=\tensl
	\textfont6=\tenbf \scriptfont6=\sevenbf \scriptscriptfont6=\fivebf
	\textfont7=\tentt
	\textfont8=\tenbi \scriptfont8=\seveni  \scriptscriptfont8=\fivemi
	\def\rm{\tenrm\fam=0 }%
	\def\it{\tenit\fam=4 }%
	\def\sl{\tensl\fam=5 }%
	\def\bf{\tenbf\fam=6 }%
	\def\tt{\tentt\fam=7 }%
	\def\ss{\tenss}%
	\def\sc{\tensc}%
	\def\bmit{\fam=8 }%
	\rm\setparameters\setbaselines\fi}


\font\twelverm=cmr12				
\font\twelvei=cmmi12				
\font\twelvesy=cmsy10	scaled\magstep1		
\font\twelveex=cmex10	scaled\magstep1		
\font\twelveit=cmti12				
\font\twelvesl=cmsl12				
\font\twelvebf=cmbx12				
\font\twelvett=cmtt12				
\font\twelvess=cmss12				
\font\twelvesc=cmcsc10	scaled\magstep1		
\font\twelvebi=cmmib10	scaled\magstep1		

\skewchar\twelvei='177
\skewchar\twelvebi='177
\skewchar\twelvesy='60

\def\twelvepoint{\ifmmode\err@badsizechange\else
	\textfont0=\twelverm \scriptfont0=\eightrm \scriptscriptfont0=\sixrm
	\textfont1=\twelvei  \scriptfont1=\eighti  \scriptscriptfont1=\sixi
	\textfont2=\twelvesy \scriptfont2=\eightsy \scriptscriptfont2=\sixsy
	\textfont3=\twelveex \scriptfont3=\tenex   \scriptscriptfont3=\tenex
	\textfont4=\twelveit \scriptfont4=\eightit \scriptscriptfont4=\sevenit
	\textfont5=\twelvesl
	\textfont6=\twelvebf \scriptfont6=\eightbf \scriptscriptfont6=\sixbf
	\textfont7=\twelvett
	\textfont8=\twelvebi \scriptfont8=\eighti  \scriptscriptfont8=\sixi
	\def\rm{\twelverm\fam=0 }%
	\def\it{\twelveit\fam=4 }%
	\def\sl{\twelvesl\fam=5 }%
	\def\bf{\twelvebf\fam=6 }%
	\def\tt{\twelvett\fam=7 }%
	\def\ss{\twelvess}%
	\def\sc{\twelvesc}%
	\def\bmit{\fam=8 }%
	\rm\setparameters\setbaselines\fi}


\font\fourteenrm=cmr12	scaled\magstep1		
\font\fourteeni=cmmi12	scaled\magstep1		
\font\fourteensy=cmsy10	scaled\magstep2		
\font\fourteenex=cmex10	scaled\magstep2		
\font\fourteenit=cmti12	scaled\magstep1		
\font\fourteensl=cmsl12	scaled\magstep1		
\font\fourteenbf=cmbx12	scaled\magstep1		
\font\fourteentt=cmtt12	scaled\magstep1		
\font\fourteenss=cmss12	scaled\magstep1		
\font\fourteensc=cmcsc10 scaled\magstep2	
\font\fourteenbi=cmmib10 scaled\magstep2	

\skewchar\fourteeni='177
\skewchar\fourteenbi='177
\skewchar\fourteensy='60

\def\fourteenpoint{\ifmmode\err@badsizechange\else
	\textfont0=\fourteenrm \scriptfont0=\tenrm \scriptscriptfont0=\sevenrm
	\textfont1=\fourteeni  \scriptfont1=\teni  \scriptscriptfont1=\seveni
	\textfont2=\fourteensy \scriptfont2=\tensy \scriptscriptfont2=\sevensy
	\textfont3=\fourteenex \scriptfont3=\tenex \scriptscriptfont3=\tenex
	\textfont4=\fourteenit \scriptfont4=\tenit \scriptscriptfont4=\sevenit
	\textfont5=\fourteensl
	\textfont6=\fourteenbf \scriptfont6=\tenbf \scriptscriptfont6=\sevenbf
	\textfont7=\fourteentt
	\textfont8=\fourteenbi \scriptfont8=\tenbi \scriptscriptfont8=\seveni
	\def\rm{\fourteenrm\fam=0 }%
	\def\it{\fourteenit\fam=4 }%
	\def\sl{\fourteensl\fam=5 }%
	\def\bf{\fourteenbf\fam=6 }%
	\def\tt{\fourteentt\fam=7}%
	\def\ss{\fourteenss}%
	\def\sc{\fourteensc}%
	\def\bmit{\fam=8 }%
	\rm\setparameters\setbaselines\fi}


\font\seventeenrm=cmr10 scaled\magstep3		


\newdimen\rp@
\newcount\@basestretchnum
\newskip\@baseskip
\newskip\headskip
\newskip\footskip


\def\setparameters{\rp@=.1em
	\headskip=24\rp@
	\footskip=\headskip
	\delimitershortfall=5\rp@
	\nulldelimiterspace=1.2\rp@
	\scriptspace=0.5\rp@
	\abovedisplayskip=10\rp@ plus3\rp@ minus5\rp@
	\belowdisplayskip=10\rp@ plus3\rp@ minus5\rp@
	\abovedisplayshortskip=5\rp@ plus2\rp@ minus4\rp@
	\belowdisplayshortskip=10\rp@ plus3\rp@ minus5\rp@
	\normallineskip=\rp@
	\lineskip=\normallineskip
	\normallineskiplimit=0pt
	\lineskiplimit=\normallineskiplimit
	\jot=3\rp@
	\setbox0=\hbox{\the\textfont3 B}\p@renwd=\wd0
	\skip\footins=12\rp@ plus3\rp@ minus3\rp@
	\skip\topins=0pt plus0pt minus0pt}


\def\setbaselines{\maxdepth=4\rp@\baselinestretch=\@basestretchnum}


\def\baselinestretch{\afterassignment\@basestretch\@basestretchnum}
\def\@basestretch{%
	\@baseskip=12\rp@ \divide\@baseskip by1000
	\normalbaselineskip=\@basestretchnum\@baseskip
	\baselineskip=\normalbaselineskip
	\bigskipamount=\the\baselineskip
		plus.25\baselineskip minus.25\baselineskip
	\medskipamount=.5\baselineskip
		plus.125\baselineskip minus.125\baselineskip
	\smallskipamount=.25\baselineskip
		plus.0625\baselineskip minus.0625\baselineskip
	\setbox\strutbox=\hbox{\vrule height.708\baselineskip
		depth.292\baselineskip width0pt }}



\def\makeheadline{\vbox to0pt{\baselinestretch=1000
	\vskip-\headskip \vskip1.5pt
	\line{\vbox to\ht\strutbox{}\the\headline}\vss}\nointerlineskip}

\def\makefootline{\baselineskip=\footskip\line{\the\footline}}

\def\big#1{{\hbox{$\left#1\vbox to8.5\rp@ {}\right.\n@space$}}}
\def\Big#1{{\hbox{$\left#1\vbox to11.5\rp@ {}\right.\n@space$}}}
\def\bigg#1{{\hbox{$\left#1\vbox to14.5\rp@ {}\right.\n@space$}}}
\def\Bigg#1{{\hbox{$\left#1\vbox to17.5\rp@ {}\right.\n@space$}}}


\mathchardef\alpha="710B
\mathchardef\beta="710C
\mathchardef\gamma="710D
\mathchardef\delta="710E
\mathchardef\epsilon="710F
\mathchardef\zeta="7110
\mathchardef\eta="7111
\mathchardef\theta="7112
\mathchardef\iota="7113
\mathchardef\kappa="7114
\mathchardef\lambda="7115
\mathchardef\mu="7116
\mathchardef\nu="7117
\mathchardef\xi="7118
\mathchardef\pi="7119
\mathchardef\rho="711A
\mathchardef\sigma="711B
\mathchardef\tau="711C
\mathchardef\upsilon="711D
\mathchardef\phi="711E
\mathchardef\chi="711F
\mathchardef\psi="7120
\mathchardef\omega="7121
\mathchardef\varepsilon="7122
\mathchardef\vartheta="7123
\mathchardef\varpi="7124
\mathchardef\varrho="7125
\mathchardef\varsigma="7126
\mathchardef\varphi="7127
\mathchardef\imath="717B
\mathchardef\jmath="717C
\mathchardef\ell="7160
\mathchardef\wp="717D
\mathchardef\partial="7140
\mathchardef\flat="715B
\mathchardef\natural="715C
\mathchardef\sharp="715D


\def\err@badsizechange{%
	\immediate\write16{--> Size change not allowed in math mode, ignored}}

\baselinestretch=1000
\tenpoint

\catcode`\@=12					
\catcode`\@=11
\expandafter\ifx\csname @iasmacros\endcsname\relax
	\global\let\@iasmacros=\par
\else	\immediate\write16{}
	\immediate\write16{Warning:}
	\immediate\write16{You have tried to input iasmacros more than once.}
	\immediate\write16{}
	\endinput
\fi
\catcode`\@=12


\def\rmb{\seventeenrm}

\def\singlespace{\baselineskip=\normalbaselineskip}
\def\halfspace{\baselineskip=1.5\normalbaselineskip}
\def\doublespace{\baselineskip=2\normalbaselineskip}


\def\AB{\bigskip\parindent=40pt
        \centerline{\bf ABSTRACT}\medskip\halfspace\narrower}
\def\AE{\bigskip\nonarrower\doublespace}
\def\nonarrower{\advance\leftskip by-\parindent
	\advance\rightskip by-\parindent}


\def\boxit#1{\vbox{\hrule\hbox{\vrule\kern3pt
	\vbox{\kern3pt#1\kern3pt}\kern3pt\vrule}\hrule}}

\def\hence{\leavevmode\hbox{\bf .\raise5.5pt\hbox{.}.} }

\def\dalemb#1#2{{\vbox{\hrule height.#2pt
	\hbox{\vrule width.#2pt height#1pt \kern#1pt \vrule width.#2pt}
	\hrule height.#2pt}}}
\def\gtorder{\mathrel{\raise.3ex\hbox{$>$}\mkern-14mu
             \lower0.6ex\hbox{$\sim$}}}
\def\ltorder{\mathrel{\raise.3ex\hbox{$<$}\mkern-14mu
             \lower0.6ex\hbox{$\sim$}}}

\newdimen\fullhsize
\newbox\leftcolumn
\def\twoup{\hoffset=-.5in \voffset=-.25in
  \hsize=4.75in \fullhsize=10in \vsize=6.9in
  \def\fullline{\hbox to\fullhsize}
  \let\lr=L
  \output={\if L\lr
        \global\setbox\leftcolumn=\columnbox\global\let\lr=R \advancepageno
      \else \doubleformat \global\let\lr=L\fi
    \ifnum\outputpenalty>-20000 \else\dosupereject\fi}
  \def\doubleformat{\shipout\vbox{
    \fullline{\box\leftcolumn\hfil\columnbox}\advancepageno}}
  \def\columnbox{\leftline{\vbox{\makeheadline\pagebody\makefootline}}}
  \tolerance=1000 }
\catcode`\@=11
\def\registered{{\ooalign
 {\hfil\raise.07ex\hbox{$^{\scriptscriptstyle\rm R}$}\hfil\crcr$^{\scriptscriptstyle
 \mathhexbox20D}$}}}
\catcode \catcode`\@=12
\twelvepoint
\doublespace
\overfullrule=0pt
{\nopagenumbers{

\rightline{~~~November, 2006}
\bigskip\bigskip
\centerline{\rmb Comments on Proposed Gravitational Modifications of}
\centerline{\rmb Schr\"odinger Dynamics and their Experimental Implications}
\medskip
\centerline{\it Stephen L. Adler}
\centerline{\bf Institute for Advanced Study}
\centerline{\bf Princeton, NJ 08540}
\medskip
\bigskip\bigskip
\leftline{\it Send correspondence to:}
\medskip
{\singlespace\leftline{Stephen L. Adler}
\leftline{Institute for Advanced Study}
\leftline{Einstein Drive, Princeton, NJ 08540}
\leftline{Phone 609-734-8051; FAX 609-924-8399; 
email adler@ias.edu}}
\bigskip\bigskip
}}
\vfill\eject
\pageno=2
\AB
We discuss aspects of gravitational modifications of Schr\"odinger 
dynamics proposed by Di\'osi and Penrose.  We consider first the 
Di\'osi--Penrose criterion for gravitationally induced state vector 
reduction, and compute the reduction time expected for a superposition 
of a uniform density 
cubical solid in two positions displaced by a small fraction of 
the cube side.   We show that the predicted effect is much smaller than 
would be observable in the proposed Marshall et al. mirror experiment.   
We then 
consider the ``Schr\"odinger--Newton'' equation for an $N$-particle 
system.  We show that in the independent particle approximation, it 
differs from the usual Hartree approximation applied to the Newtonian 
potential by self-interaction terms, which do not have a consistent 
Born rule interpretation.  This raises doubts about the use of the 
Schr\"odinger--Newton equation to calculate gravitational effects on molecular interference experiments. 
When the effects of Newtonian gravitation on 
molecular diffraction are calculated using the standard many-body  
Schr\"odinger equation, no washing out of the interference 
pattern is predicted.  
\AE
\bigskip\bigskip
\vfill\eject
\pageno=3
\centerline{{\bf 1.~~Introduction}}
There is now considerable interest in mounting experiments to search for,  
and/or to place limits on, 
possible modifications of Schr\"odinger dynamics.  We focus in this paper 
on conjectured gravitational modifications of the Schr\"odinger equation 
associated with the work of Di\'osi [1], Penrose [2] and their 
collaborators.  These authors have proposed a gravitationally based 
criterion, which we refer to as the Di\'osi--Penrose (DP) criterion, for 
predicting when a superposition of two spatially displaced states of the 
same object will reduce to either one state or the other.  In Sec. 2 
we briefly review the DP criterion and its theoretical motivations,   
including the gravitationally driven stochastic equation formulated by 
Di\'osi [1]. In Sec. 3 we evaluate  the  DP effect for a uniform 
cube displaced 
by a small fraction of its side, and show that the predicted rate of 
gravitational state vector reduction is too small to be observed in 
the proposed Marshall et al. [3] mirror superposition experiment.  A 
different, non-gravitational criterion based on displacement of the 
center-of-mass wave packet, will however be tested by the Marshall et al. 
proposal. 

Di\'osi [4] and Penrose [2] have also proposed a nonlinear equation, 
called 
the ``Schr\"od-\hfill\break
inger--Newton''(SN) equation, for including non-stochastic 
effects of gravitation on quantum evolution.  In Sec. 4 we review the 
SN equation, give its specialization in the independent particle approximation, 
and contrast this with the standard Hartree approximation as applied to the 
inter-particle Newtonian potential.  We show that the two differ by 
a particle self-interaction term, which is not included in the standard 
Hartree approximation to Newtonian dynamics, and which does not have 
a consistent probabilistic interpretation within the framework of the 
Born rule. 
 Salzman and Carlip [5], motivated by searching for distinctive features 
of non-quantized gravitation,  have recently argued that the 
SN equation implies potentially observable effects in molecular diffraction 
experiments.  In Sec. 5 we consider gravitational effects on 
molecular diffraction  in standard many-body quantum theory 
as applied to the inter-particle Newtonian potential, which omits  
the suspect self-interaction effect of the SN equation.  We show 
(without invoking the Hartree approximation)  that there 
is a complete decoupling of gravitational effects 
from the center-of-mass motion of the  molecule,  and thus 
no reduction in visibility of molecular interference fringes is predicted.  

\bigskip
\centerline{{\bf 2.~~The Di\'osi--Penrose (DP) criterion and Di\'osi's 
stochastic Schr\"odinger equation}}
\bigskip
Di\'osi [1] proposed that there is a ``universal gravitational white 
noise'', represented by a stochastic  term $\phi(r,t)$ in the 
gravitational potential (where $r$ is the coordinate three-vector).  
Denoting the stochastic expectation by 
$E[...]$, this fluctuating part of the gravitational potential is 
assumed to obey  
$$\eqalign{
E[\phi (r,t)]=&0~~~,\cr
E[\phi(r,t)\phi(r^{\prime},t)]=&\hbar G |r-r^{\prime}|^{-1}
\delta(t-t^{\prime})~~~,\cr
}\eqno(1)$$
with $G$ the Newton gravitational constant.  
Including $\phi$ in the Schr\"odinger equation, Di\'osi is led 
to a stochastic dynamics 
$$i\hbar \dot \psi(t)=\left(H + \int d^3r \phi(r,t) f(r)\right)\psi(t)~~~,
\eqno(2a)$$ 
with $H$ the usual Hamiltonian and $f(r)$ the local 
mass density operator. 
This in turn implies that the stochastic expectation density matrix 
$\rho(t)=E[\psi(t) \psi(t)^{\dagger}]$  obeys the dynamical  
equation
 $$\dot \rho(t) = {-i \over \hbar} [H, \rho(t)] 
-{G \over 2 \hbar} \int\int  {d^3r d^3r^{\prime} \over |r-r^{\prime}| }
[f(r),[f(r^{\prime}),\rho(t)]]~~~.\eqno(2b)$$

Letting $X$ denote the system coordinates, and $f(r|X)$ the mass 
density at $r$ for the system configuration $X$, Eq.~(2b) implies that 
the off-diagonal matrix element $\langle X|\rho(t)|X^{\prime}\rangle$ 
damps with a characteristic time $\tau_d(X,X^{\prime})$ given by 
$$\tau_d(X,X^{\prime})^{-1}=
{G\over 2 \hbar} \int\int d^3r d^3r^{\prime}
 {[f(r|X)-f(r|X^{\prime})][f(r^{\prime}|X) 
-f(r^{\prime}|X^{\prime})] \over |r-r^{\prime}| }   
~~~.\eqno(3)$$
(Note that Eq.~(12) of Di\'osi's paper where $\tau_d$ is defined  contains an algebraic error, 
and should read as in Eq.~(3) above, which is what one gets when one 
takes the off-diagonal matrix element of Di\'osi's Eq.~(11).  This error 
was noted some time ago by Anandan [6].)  Although the 
density matrix evolution of Eq.~(3) leads to exponential damping in time 
of the off-diagonal 
density matrix element $\langle X|\rho(t)|X^{\prime}\rangle$,
  the stochastic Schr\"odinger 
equation of Eq.~(2a) does not lead to state vector reduction, since 
 an initial 
superposition of configurations $X$ and $X^{\prime}$ does not evolve 
to just one of the two alternatives.  However, a non-linear variant 
of Eq.~(2a), constructed according to the continuous spontaneous 
localization scheme reviewed by 
Bassi and Ghirardi [7] and Pearle [7], does lead 
to state vector reduction, with the stochastic expectation 
density matrix also obeying the evolution equation of Eq.~(2b).

Penrose [2] has also proposed a role for 
gravitation in state vector reduction, based on the observation that 
when a macroscopic mass distribution is moved significantly, the 
spacetime geometry is changed.  Since standard quantum theory does not 
permit the description of coherent superpositions of states 
constructed on two different background geometries, Penrose argues that 
in a correct theory that merges spacetime geometry with quantum theory, 
such coherences must decay.  He thus arrives at a criterion which states 
that a coherent superposition of matter density distributions 
$\rho(x)$ and $\rho^{\prime}(x)$ should reduce to one or the other in 
a characteristic time $\tau_d^{-1}=\Delta/\hbar$, 
with $\Delta$ given by 
$$\Delta= G \int\int d^3r d^3r^{\prime} {[\rho(r)-\rho^{\prime}(r)] 
[\rho(r^{\prime})-\rho^{\prime}(r^{\prime})] \over 
|r-r^{\prime}|}~~~.\eqno(4)$$ 
(In his papers, Penrose uses the notation $x,y$ for what we have termed 
$r,r^{\prime}$, and his 2000 paper [2] giving Eq.~(4) differs by a factor 
of $4\pi$ from his 1996 paper [2].  We will follow the later version, 
and will  reserve the designation $x,y,z$ for the Cartesian 
components of $r$.) 
Apart from obvious differences in notation, and an extra numerical factor 
of $2$, Penrose's criterion of Eq.~(4) is the same as Di\'osi's 
criterion of Eq.~(3), and we shall refer to the two collectively 
as the Di\'osi--Penrose (DP) criterion.  

Because Eq.~(4) diverges for point particles, the effect predicted 
depends on the radius assigned to the elementary mass distributions.  
Moreover, the density matrix evolution of Eq.~(2b) predicts energy 
non-conservation, which as discussed by Ghirardi, Grassi, and Rimini [8],
disagrees with experimental bounds unless the point particle 
mass distributions are smeared considerably more than originally 
envisaged by Di\'osi.    Rather than adding a smearing radius as 
an additional parameter of the model, we note that 
for any smearing radius greater than 
a typical interatomic distance of $10^{-8}$ cm, the mass 
distribution becomes effectively uniform.  Motivated by this, 
we shall assume a homogeneous mass distribution 
in applying the DP criterion. 
\bigskip
 
\centerline{\bf 3.~~Magnitude of the DP estimator in the 
Marshall et al. mirror experiment}
\bigskip
Continuing with Eq.~(4), with mass distributions assumed homogeneous, 
let us consider the specific geometry of the 
Marshall et al. [3] proposal, in which a cubical mirror with side 
$S=10^{-3}$ cm is put into a superposition of two states displaced 
parallel to a side of the cube by $d= 
10^{-11}$ cm.  Since the displacement $d$ is a small fraction of the 
mirror dimension $S$, we follow Di\'osi [9] and Geszti [10] and expand 
Eq.~(4) to leading, quadratic order in $d$.  
Writing 
$$\eqalign{
\rho(r)=&\rho_0\theta(S-x)\theta(x)\theta(S-y)\theta(y)
\theta(S-z)\theta(z) ~~~,\cr      
\rho^{\prime}(r)=&\rho_0\theta(S-x)\theta(x)\theta(S-y)\theta(y)
\theta(S-z-d)\theta(z+d) ~~~,\cr      
}\eqno(5a)$$ 
with $\theta(x)$ the standard step function that jumps from 0 to 1 
at $x=0$, 
we have 
$$\rho(r)-\rho^{\prime}(r)=\rho_0\theta(S-x)\theta(x)\theta(S-y)\theta(y)
[\theta(S-z)\theta(z)-\theta(S-z-d)\theta(z+d)] ~~~.\eqno(5b)$$
Substituting 
$$\eqalign{
\theta(S-z-d)\simeq& \theta(S-z) -d\delta(S-z)~~~\cr 
\theta(z+d) \simeq& \theta(z) + d \delta(z) ~~~~,\cr
}\eqno(5c)$$
we find 
$$\rho(r)-\rho^{\prime}(r)\simeq d\rho_0\theta(S-x)\theta(x)\theta(S-y)\theta(y) 
[-\delta(z)+\delta(S-z)] ~~~,\eqno(5d)$$
with a similar expression with all coordinates replaced by primed 
coordinates.  Thus Eq.~(4) becomes 
$$\Delta = G d^2 \rho_0^2 I_1~~~,\eqno(6a)$$
with $I_1$ given by 
$$\eqalign{
I_1=&\int \int d^3r d^3 r^{\prime} \theta(S-x)\theta(x) \theta(S-y) 
\theta(y) \theta(S-x^{\prime}) \theta(x^{\prime}) \theta(S-y^{\prime}) 
\theta(y^{\prime})\cr
\times& [-\delta(z)+\delta(S-z)] [-\delta(z^{\prime})+\delta(S-z^{\prime})]
 [(x-x^{\prime})^2+(y-y^{\prime})^2+(z-z^{\prime})^2]^{-1/2}~~~.\cr
 }\eqno(6b)$$

Using the delta functions to eliminate the $z,z^{\prime}$ integrals, 
imposing the theta function constraints on the 
$x,y,x^{\prime},y^{\prime}$ integrals and 
and scaling out the cube side $S$,  we get 
finally 
$$\Delta=2 G d^2 S^3 \rho_0^2 I~~~, \eqno(7a)$$
with I the dimensionless integral given by 
$$I=\int_0^1 dx \int_0^1 dy \int_0^1 dx^{\prime} \int_0^1 dy^{\prime} 
\left( {1\over [(x-x^{\prime})^2 + (y-y^{\prime})^2]^{1/2} } 
-   {1\over [(x-x^{\prime})^2 + (y-y^{\prime})^2+1]^{1/2} } \right)~~~.
\eqno(7b)$$ 
The quadruple integral $I$ can be simplified by transforming to sum and difference 
variables $\eta_x=x-x^{\prime}~,~~ \sigma_x=x+x^{\prime}$, etc., giving 
the double integral form 
$$\eqalign{
I=&4\int_0^1 d\eta_x \int_0^1 d\eta_y (1-\eta_x) (1-\eta_y) 
\left( {1\over [\eta_x^2+\eta_y^2]^{1/2} }   
- {1\over [\eta_x^2+\eta_y^2+1]^{1/2} }  \right) \cr  
=&2\pi/3\simeq 2.0944 ~~~.\cr
}\eqno(7c)$$
The evaluation of the integral on the first line of Eq.~(7c) 
 was done using 
Mathematica\registered; as a check we also used Mathematica\registered\  to numerically 
evaluate  the 
quadruple integral of Eq.~(7b), giving the same result.  

Putting everything together, we have
$$\Delta=(4\pi/3) G d^2 S^3 \rho_0^2~~~,\eqno(8a)$$
which with $d=10^{-11}$ cm, $S=10^{-3}$ cm, and $\rho_0S^3=5 \times 
10^{-12}$ kg gives 
$$\eqalign{
\Delta=&2.2\times 10^{-20} \hbar c~ {\rm cm}^{-1}~~~,\cr 
\tau_d=&\hbar/\Delta=1.5\times 10^9  {\rm s}~~~.\cr
}\eqno(8b)$$
Hence, the characteristic time for gravitational effects on the superposed 
cube wave function, according to the DP criterion, is much longer than 
the observation time interval of the Marshall et al. proposal, which is given 
in terms of the mirror oscillation angular frequency $\omega_m$ by 
$2\pi/\omega_m=2\times 10^{-3} {\rm s}$.  

Thus, the Marshall et al. 
proposal, even it achieves the sought-for sensitivity, will not 
confront the DP proposal for state vector reduction, when interpreted  
using homogeneous mass distributions.  We emphasize at this point that  
the Marshall et al. paper does not suggest that it will test gravitationally 
induced reduction models (although citation of the Penrose papers [2] 
in the Marshall et al. proposal might lead  readers to conclude 
otherwise). The mirror experiment proposal suggests 
a different, non-gravitational, criterion for state vector reduction, that 
superpositions 
reduce when an object is displaced by more than the width of the 
center-of-mass wave packet, and this condition is met by the 
proposed experiment.   
The purpose of the exercise we have just gone through has been, first of 
all, to get the explicit formula for the DP criterion in the context of 
the mirror experiment, and secondly, 
to demonstrate that the DP criterion and 
the center-of-mass displacement criterion can make very different predictions. 
For completeness, we note that the mirror experiment may also be sensitive  
to other types of spontaneous localization models, if the stochasticity 
magnitude  is taken large enough to give state vector reduction in 
latent image formation, as discussed in Adler [11] (which draws on 
earlier analyses of the mirror experiment in [12]). 
\bigskip

\centerline{\bf 4.~~The ``Schr\"odinger-Newton'' (SN) equation in the 
independent particle} 
\centerline{\bf approximation versus the Hartree approximation}
\bigskip
As an attempt to incorporate quantized matter into a purely classical 
theory of gravitation, M\o ller  [13] and Rosenfeld [14] have suggested that the 
source term in the classical Einstein equation be taken as the expectation  
$\langle \psi | T_{\mu \nu} | \psi \rangle$ of the energy momentum operator 
$ T_{\mu \nu} $ in the quantum state $|\psi\rangle$.  As a nonrelativistic  
realization of this idea, Di\'osi [4] and Penrose [2] have proposed 
what has come to be called the ``Schr\"odinger--Newton'' equation, in 
which a quantum many-body system of $N$ particles moves in a gravitational 
potential given by the quantum expectation of the operator Newtonian 
potential.  Following the exposition of Di\'osi [4], the many-body 
equation for particles of masses $m_1,...,m_N$ is taken as 
$$
i\hbar \partial \psi(X,t)/\partial t 
=\left(-\sum_{r=1}^N {\hbar^2 \over 2 m_r} {\partial^2 \over \partial x_r^2} 
+ \sum_{r,s=1}^N V_{rs}(x_r-x_s)   
+ \sum_{s=1}^N m_s \phi(x_s,t) \right) \psi(X,t)~~~.\eqno(9a)$$
Here $V_{rs}$ is a non-gravitational interaction potential, which we  
shall ignore for the present discussion, $X=(x_1,x_2,...,x_N)$ denotes the 
spatial coordinates of the $N$ particles, and $\phi(x)$ is the Newtonian 
gravitational potential obtained from the nonrelativistic specialization 
of the M\o ller-Rosenfeld equation.  In other words, $\phi$ is obtained  
by solving 
$$\nabla^2 \phi(x,t) = 4\pi G 
\int d^{3N}X^{\prime} |\psi(X^{\prime},t)|^2 \sum_{u=1}^N m_u 
\delta^{(3)}(x-x_u^{\prime})~~~,\eqno(9b)$$ 
where $X^{\prime}=(x_1^{\prime},...,x_N^{\prime})$.  Inverting Eq.~(9b) 
and substituting into Eq.~(9a), we get the Schr\"odinger--Newton equation 
$$\eqalign{
i\hbar \partial \psi(X,t)/\partial t 
=&\left(-\sum_{r=1}^N {\hbar^2 \over 2 m_r} {\partial^2 \over \partial x_r^2} 
+ \sum_{r,s=1}^N V_{rs}(x_r-x_s)\right. \cr  
 -&\left. G\sum_{u,s=1}^N \int  d^{3N}X^{\prime}  {m_u m_s \over |x_s-x_u^{\prime}|}
 |\psi(X^{\prime},t)|^2 \right)  \psi(X,t)  ~~~.\cr
 }\eqno(9c)$$
For a single particle of mass $m$, this reduces to a  Schr\"odinger equation 
with a nonlinear and nonlocal interaction term, 
$$ 
i\hbar \partial \psi(x,t)/\partial t 
=-{\hbar^2 \nabla^2 \over 2m} \psi(x,t) 
-Gm^2 \int d^3x^{\prime} {|\psi(x^{\prime},t)|^2 \over |x-x^{\prime}|} 
\psi(x,t)~~~.\eqno(9d)$$

Let us now specialize Eq.~(9c) to the case when the non-gravitational 
interaction $V_{rs}$ vanishes, so that it becomes 
$$i\hbar \partial \psi(X,t)/\partial t 
=\left(-\sum_{r=1}^N {\hbar^2 \over 2 m_r} {\partial^2 \over \partial x_r^2} 
- G\sum_{u,s=1}^N \int  d^{3N}X^{\prime}  {m_u m_s \over |x_s-x_u^{\prime}|}
 |\psi(X^{\prime},t)|^2 \right)  \psi(X,t)  ~~~. \eqno(10)$$
We wish to study the form taken by Eq.~(10) when we make an 
independent particle Ansatz, 
$$\psi(X,t) = \prod_{r=1}^N \psi_r(x_r,t)~~~,\eqno(11a)$$
with each single particle wave function $\psi_r$ normalized to unity,
$$\int d^3x_r |\psi_r(x_r,t)|^2 =1~~~.\eqno(11b)$$
Substituting Eq.~(11a) into Eq.~(10) and using Eq.~(11b), and dividing 
by $\psi(X,t)$, we get 
$$\sum_{s=1}^N F(x_s,t)/\psi_s(x_s,t) =0~~~,\eqno(12a)$$
with 
$$F(x_s,t)=-i\hbar \partial \psi_s(x_s,t)/\partial t 
 - {\hbar^2 \over 2 m_s} {\partial^2 \over \partial x_s^2} \psi_s(x_s,t) 
- G\sum_{u=1}^N \int  d^3x_u^{\prime}  {m_u m_s \over |x_s-x_u^{\prime}|}
 |\psi_u(x_u^{\prime},t)|^2   \psi_s(x_s,t)  ~~~. \eqno(12b)$$
Since the different terms in Eq.~(12a) involve independent variables 
$x_s$, the usual separation of variables argument implies that each must 
be a constant, 
$$ F(x_s,t))/\psi_s(x_s,t)=c_s~~~,\eqno(13a)$$ 
with the constants $c_s$ summing to zero, 
$$\sum_{s=1}^N c_s=0~~~.\eqno(13b)$$
However, if we introduce new single-particle wave functions 
$\hat \psi_s(x_s,t)$ through 
$$\psi_s(x_s,t)=\exp(i c_s t/\hbar)\hat\psi_s(x_s,t) ~~~,\eqno(13c)$$
then by virtue of Eq.~(13b), we have  
$$\prod_{r=1}^N \psi_r(x_r,t)=\prod_{r=1}^N\hat \psi_r(x_r,t)~~~,\eqno(13d)$$
and Eq.~(13a) becomes $\hat F(x_s,t)=0$, where $\hat F(x_s,t)$ is 
obtained from $F(x_s,t)$ of Eq.~(12b) by replacing $\psi_s$ by $\hat \psi_s$. 
We thus conclude that there is no loss of generality in taking the 
separation constants $c_s$ all as zero, and the single-particle equation 
as $F(x_s,t)=0$, that is 
$$i\hbar \partial \psi_s(x_s,t)/\partial t=      
 - {\hbar^2 \over 2 m_s} {\partial^2 \over \partial x_s^2} \psi_s(x_s,t) 
- G\sum_{u=1}^N \int  d^3x_u^{\prime}  {m_u m_s \over |x_s-x_u^{\prime}|}
 |\psi_u(x_u^{\prime},t)|^2   \psi_s(x_s,t)  ~~~. \eqno(14)$$

Equation (14) has an almost familiar look; it has the same structure as 
the time-dependent single particle equation that one would get by treating 
the Newtonian inter-particle potential in the Hartree approximation, 
except that it includes a self-interaction term coming from the $u=s$ 
term in the summation, 
$$
- G \int  d^3x_s^{\prime}  {m_s^2  \over |x_s-x_s^{\prime}|}
 |\psi_s(x_s^{\prime},t)|^2   \psi_s(x_s,t)  ~~~. \eqno(15a)$$
Such self-interaction terms of a single particle never appear in the 
Hartree equation, and do not have a consistent interpretation within 
the Born rule interpretation of quantum theory.  A term with 
$u \not= s$ in the potential energy of Eq.~(14), 
$$
- G \int d^3x_u^{\prime}  {m_u m_s \over |x_s-x_u^{\prime}|}
 |\psi_u(x_u^{\prime},t)|^2   \psi_s(x_s,t)  ~~~, \eqno(15b)$$
has the interpretation that the gravitational potential felt 
by particle $s$ at coordinate $x_s$, as a result of the presence of 
particle $u$ at $x_u^{\prime}$, is the Newtonian potential 
$-Gm_um_s/  |x_s-x_u^{\prime}|$ weighted by the probability 
$ |\psi_u(x_u^{\prime},t)|^2 $ of finding particle $u$ at $x_u^{\prime}$.  
However, this interpretation does not extend to the case $u=s$, since 
when particle $s$ is at $x_s$, the probability of simultaneously finding 
it at $x_s^{\prime}$ is zero!  In terms of projectors, in the 
case $u\not= s$ we have that $P_u(x_u^{\prime}) P_s(x_s) $ gives a nonzero projector 
for finding particle $s$ at $x_s$, and particle $u$ simultaneously 
at $x_u^{\prime}$.  However, in the case $u=s$ we have $P_s(x_s^{\prime})
P_s(x_s)=0$ for all $x_s^{\prime} \not= x_s$.

We conclude from this analysis that the Schr\"odinger-Newton equation 
does not give a consistent interpretation of the mutual gravitational 
interactions within a single  system of particles.  It can, however, be used 
to calculate the gravitational effect of one group of 
particles on a disjoint group of particles (say, of the sun on a planet), 
since then the problematic  self-interaction terms are not present.  

\vfill\eject
\centerline{\bf 5.~~Gravitational effects on molecular scattering in  
standard}
\centerline{\bf  many-body quantum theory}
\bigskip
In a recent archive posting, Salzman and Carlip [5]  studied 
the single particle case, Eq.~(9d), of the SN equation, and based on 
this suggested  
that there may be  significant nonlinear gravitational effects in potentially 
observable situations, such as molecular interferometry experiments.  
However, the single particle case of the SN equation consists of a 
self-interaction term which, as we observed in 
the preceding section,  does not appear in 
the standard Hartree approximation, and which does not have a Born rule 
interpretation. This makes it problematic, we believe, to apply the 
SN equation to the mutual gravitational interactions within a system 
of atoms, as needed, for example, to discuss gravitational effects in 
molecular diffraction.

There is a standard way of treating gravitational effects on large 
molecules, within conventional many-body theory (without use of the Hartree
approximation), which leads to a  different conclusion from that reached   
in [5].  One simply 
includes in the interaction term 
$$ \sum_{r,s=1}^N V_{rs}(x_r-x_s) ~~~\eqno(16a)$$  
of Eq.~(9a) a Newtonian gravitational potential term 
$$-{1\over 2} \sum_{r \not= s} {Gm_rm_s\over |x_r-x_s|} ~~~,\eqno(16b)$$
in analogy with the usual treatment of the inter-particle 
Coulomb potential. 
Since both Eq.~(16a) (which includes the Coulomb force terms) and 
Eq.~(16b) (which is the gravitational perturbation) depend only on the 
relative coordinates $x_r-x_s$, they do not appear in the equation for 
the center-of-mass wave function of the molecule.  The center of mass will 
thus obey a free-particle Schr\"odinger equation, subject to external 
potentials (such as diffraction gratings) acting on the molecule.  
Therefore, one expects no significant 
 effect on the molecular interference 
pattern from the mutual gravitational interactions of the molecular   
constituents.  Such gravitational perturbations will very slightly 
change the shape and energy levels of the molecule, but will not exert 
an influence on its center-of-mass motion, other than 
(when relativistic effects are taken into account) through their 
small  modification of the mass of the molecule. 

We conclude with a question that suggests further work.  
As noted above, the SN equation arises from 
applying the M\o ller--Rosenfeld semiclassical equation to the Newtonian 
interaction of a many-particle system.  Do the problems that we have 
encountered indicate that a semiclassical approach to gravitation is 
inconsistent, and hence that gravity must be quantized [15]?  
Or do they only 
indicate that a modification of the M\o ller--Rosenfeld and SN approach 
should be sought, which will make possible a consistent semiclassical 
theory of gravitational effects?

\bigskip
\centerline{\bf Acknowledgments}
This work was supported in part by the Department of Energy under
Grant \#DE--FG02--90ER40542.  I wish to thank Angelo Bassi, 
Steven Carlip, and Lajos 
Di\'osi for stimulating discussions at the DICE 2006 conference organized by 
Hans-Thomas Elze, as well as for email 
comments on the initial  draft of this paper.    
\vfill\eject
\centerline{\bf References}
\bigskip
\noindent
[1]  Di\'osi L 1987 {\it Phys. Lett. A} {\bf 120} 377;  
Di\'osi L 1989 {\it Phys. Rev. A} {\bf 40} 1165;  
Di\'osi L 2005 {\it Braz. J. Phys.} {\bf 35} 260 
arXiv:quant-ph/0412154; see also [9]  
\hfill\break
\bigskip 
\noindent
[2] Penrose R 1996 {\it Gen. Rel. Grav.} {\bf 28} 581; Penrose R 
1998 {\it Phil Trans. R. Soc. Lond. A} {\bf 356} 1927; Penrose 
R 2000 Wavefunction Collapse as a Real Gravitational Effect 
{\it Mathematical Physics 2000} ed A Fokas et al. (London: Imperial 
College)\hfill\break 
\bigskip
\noindent
[3]  Marshall W, Simon C, Penrose R and Bouwmeester D 2003 {\it Phys. Rev. Lett.} 
{\bf 91} 130401 \hfill\break
\bigskip
\noindent
[4]  Di\'osi L 1984 {\it Phys. Lett. A} {\bf 105} 199; 
see also Di\'osi L and Luk\'acs B 1987 {\it Ann. Phys. Leipzig} 
{\bf 44} 488 \hfill\break
\bigskip
\noindent
[5]  Salzman P J and Carlip S 2006 ``A possible experimental test 
of quantized gravity'' arXiv:gr-qc/0606120 \hfill\break
\bigskip
\noindent
[6]  Anandan J 1998  communication to L Di\'osi; see Di\'osi L 2005  
{\it Braz. J. Phys.} {\it 35} 260 arXiv:quant-ph/0412154\hfill\break
\bigskip
\noindent
[7] Bassi A and Ghirardi G C 2003 Dynamical reduction models 
{\it Phys. Rep.} {\bf 379} 257; Pearle P 1999 Collapse Models {\it Open 
Systems and Measurements in Relativistic Quantum Field Theory (Lecture 
Notes in Physics vol 526)} ed H-P Breuer and F Petruccione (Berlin: 
Springer) \hfill\break 
\bigskip
\noindent
[8] Ghirardi G C, Grassi R and Rimini A 1990 {\it Phys. Rev. A} 
{\bf 42} 1057 \hfill\break
\bigskip
\noindent
[9] Di\'osi L 2006 ``Notes on Certain Newton Gravity Mechanisms of 
Wave Function Localization and Decoherence'' {\it J. Phys. A: Math Gen} 
in press \hfill\break
\bigskip
\noindent
[10] Geszti T 2004 {\it Phys. Rev. A} {\bf 69} 032110 \hfill\break
\bigskip
\noindent
[11]  Adler S L 2006 ``Lower and Upper Bounds on CSL Parameters from 
Latent Image Formation and IGM Heating'' arXiv:quant-ph/0605072 
{\it J. Phys. A: Math Gen} in press \hfill\break 
\bigskip
\noindent
[12] Bassi A, Ippoliti E and Adler S L 2005 {\it Phys. Rev. Lett.} 
{\bf 94} 030401; Adler S L, Bassi A and Ippoliti E 2005 
{\it J. Phys. A: Math. Gen.} {\bf 38} 2715; Bern\'ad J Z, 
Di\'osi L and Geszti T 2006 ``Quest for quantum superpositions of a 
mirror: high and moderately low temperatures'' arXiv:quant-ph/0604157
\hfill\break
\bigskip
\noindent 
[13] M\o ller C 1962 {\it Les Th\'eories Relativistes de la 
Gravitation} Colloques Internationaux CNRS 91 ed A Lichnerowicz and 
M-A Tonnelat (Paris: CNRS) \hfill\break
\bigskip
\noindent
[14] Rosenfeld L 1963 {\it Nucl. Phys.} {\bf 40} 353 \hfill\break
\bigskip
\noindent
[15] Page D N and Geilker C D 1981 {\it Phys. Rev. Lett.} 
{\bf 47} 979; Ballentine L E 1982 {\it Phys. Rev. Lett.}  {\bf 48} 
522; Hawkins B 1982 {\it Phys. Rev. Lett.} {\bf 48} 520; 
Unruh W G 1984 Steps towards a Quantum Theory of Gravity 
{\it Quantum Theory of Gravity: Essays in honor of the 60th birthday 
of Bryce S. De Witt} ed S M Christensen (London: Hilger) \hfill\break

\bigskip
\noindent
\bigskip
\noindent
\bigskip
\noindent
\bigskip
\noindent
\bigskip
\noindent
\bigskip
\noindent
\vfill
\eject
\bigskip
\bye